\documentclass[twocolumn,secnumarabic,amssymb, amsmath, nobibnotes, aps,prb,superscriptaddress,preprintnumbers]{revtex4-2}

\usepackage{graphicx}
\usepackage{dcolumn}
\usepackage{bm}
\usepackage{color}
\usepackage{xcolor}
\usepackage{comment}
\usepackage{float}

\begin{document}

\title{Carrier-doping effect and anomalous transport properties in Ni-doped CeCoIn$_{\bf {5}}$ investigated by Hall resistivity measurements}%

\author{Ryosuke~Koizumi}
\email[]{Contact author: 25nd103a@vc.ibaraki.ac.jp}
\author{Hayao~Fujimoto}
\author{Teppei~Takahashi}
\author{Azumi~Yashiro}
\author{Haruna~Kawakami}
\author{Kazuki~Ishii}
\author{Hinako~Kosaka}
\author{Takeshi~Hasegawa}
\affiliation{College of Science, Ibaraki University, Mito, Ibaraki 310-8512, Japan}
\author{Yusei~Shimizu}
\altaffiliation{Present address: Institute for Solid State Physics, The University of Tokyo, Kashiwa, Chiba 277-8581, Japan}
\author{Ai~Nakamura}
\author{Dai~Aoki}
\affiliation{Institute for Materials Research, Tohoku University, Oarai, Ibaraki 311-1313, Japan}
\author{Kenichi~Tenya}
\affiliation{Faculty of Education, Shinshu University, Nagano 380-8544, Japan}
\author{Makoto~Yokoyama}
\email[]{Contact author: makoto.yokoyama.sci@vc.ibaraki.ac.jp}
\affiliation{College of Science, Ibaraki University, Mito, Ibaraki 310-8512, Japan}
\affiliation{Research and Education Center for Atomic Sciences, Ibaraki University, Tokai, Ibaraki 319-1106, Japan}
\date{\today}
             
\begin{abstract}
We investigated the effects of Ni doping on carrier density and anomalous electrical transport properties in CeCo$_{1-x}$Ni$_x$In$_5$ ($x \leq 0.3$) by performing Hall resistivity measurements. The carrier density, estimated from the Hall coefficient $R_{\rm H}$ at a temperature of 0.5 K in high magnetic fields, increases linearly with $x$, indicating that the doped Ni ions act as electron dopants. In CeCoIn$_5$, the magnitude of $-R_{\rm H}$ is strongly enhanced at magnetic fields near the superconducting upper critical field $H_{c2}$ and in the low-field region above the superconducting transition temperature $T_c$. However, these anomalies are found to be significantly suppressed by Ni doping. Possible origins of this suppression in $-R_{\rm H}$ are discussed.
\end{abstract}

\maketitle
\section{Introduction}
Hall resistivity measurements are a powerful technique for investigating quantum critical phenomena in various itinerant electron systems \cite{RevModPhys.82.1539,Nair01102012}. The quantum critical phenomena arising from antiferromagnetic (AFM) spin fluctuations are typically induced by continuously suppressing the AFM order through the application of magnetic fields, pressure, or chemical substitutions. The resulting critical condition, where the AFM transition temperature approaches zero ($T_N \to 0$), is referred to as the AFM quantum critical point (QCP). In itinerant AFM compounds exhibiting quantum criticality, such as YbRh$_2$Si$_2$ and Cr, the Hall resistivity measurements have revealed changes in the carrier density across the QCP \cite{paschen2004hall,doi:10.1073/pnas.1009202107,PhysRevLett.92.187201,doi:10.1073/pnas.1005036107}. Thus, it is expected that the carrier density near the AFM QCP provides valuable insights into the origin of QCP formation, as many theoretical models have been proposed to explain it \cite{si2001locally,P.Coleman_2001,PhysRevB.69.035111,paschen2021quantum}.

The heavy-fermion superconductor CeCoIn$_5$ exhibits quantum critical phenomena arising from itinerant quasiparticles in various physical quantities, including Hall resistivity \cite{doi:10.1143/JPSJ.73.5,PhysRevB.70.035113,doi:10.1143/JPSJ.76.024703,PhysRevLett.98.057001,PhysRevB.87.045102}. CeCoIn$_5$ crystallizes in the HoCoGa$_5$-type tetragonal structure (see Fig.~1 inset) and exhibits superconductivity with a $d_{x^2-y^2}$-type gap symmetry below the superconducting (SC) transition temperature $T_c = 2.3~{\rm K}$ \cite{Petrovic_2001,PhysRevLett.87.057002,PhysRevLett.100.177001,PhysRevLett.104.037002}. Anomalous SC properties have been reported thus far, and most of them likely originate from AFM spin correlations and quantum criticality \cite{PhysRevLett.91.246405,PhysRevLett.91.257001,PhysRevLett.111.107003}. For instance, a strong Pauli paramagnetic effect enhanced under an applied magnetic field $H$ leads to a first-order transition at the SC upper critical field $H_{c2}$, and an additional SC phase, termed the $Q$ phase, emerges just below $H_{c2}$ for $H$ applied along the tetragonal $c$ plane \cite{PhysRevB.65.180504,PhysRevLett.91.187004,PhysRevLett.94.047602,PhysRevLett.98.036402,doi:10.1126/science.1161818,Aperis_2008,PhysRevLett.102.207004,doi:10.1143/JPSJ.77.063705,doi:10.1143/JPSJ.78.114715,PhysRevB.107.L220505}. The inelastic neutron scattering experiments have revealed that a resonance excitation involving the tetragonal $c$-axis spin polarization develops in the SC state \cite{PhysRevLett.100.087001,PhysRevLett.115.037001,10.1038/ncomms12774,PhysRevLett.119.187002}.

Various ion-substituted systems have been investigated to gain a deeper understanding of these anomalous SC properties. When ions with atomic numbers one less than those of the constituent atoms in CeCoIn$_5$ are substituted (e.g., Cd, Hg, and Zn at the In site), increasing the substitution level suppresses the SC order and simultaneously induces AFM orders \cite{PhysRevLett.97.056404,PhysRevB.76.052401,PhysRevLett.99.146402,seo2014disorder,PhysRevLett.121.037003,doi:10.7566/JPSJ.83.033706,PhysRevB.92.184509,PhysRevB.95.224425,doi:10.7566/JPSJ.87.034702,PhysRevB.104.085106,PhysRevB.105.054515,doi:10.1073/pnas.2209549119,PhysRevB.111.104510}. In contrast, when ions with atomic numbers one higher than those of the constituent atoms are substituted (e.g., Sn at the In site and Pt and Ni at the Co site), the ion doping monotonically suppresses the SC order and subsequently leads to the non-Fermi liquid (NFL) state without inducing magnetic order \cite{PhysRevLett.94.047001,PhysRevB.73.245109,PhysRevLett.105.126401,doi:10.7566/JPSJ.85.094713,PhysRevB.99.054506,PhysRevB.106.235152,PhysRevMaterials.8.L081801}. These systematic characteristics suggest that carrier doping is key for tuning the SC, AFM, and NFL states realized in CeCoIn$_5$ and its doped alloys \cite{PhysRevLett.109.186402,PhysRevB.97.045134,PhysRevB.105.115121,doi:10.1126/science.aaz4566,PhysRevB.109.125147}. Therefore, quantitative information on changes in carrier density is essential for elucidating properties of these states.

In the mixed compound CeCo$_{1-x}$Ni$_x$In$_5$, the Ni doping continuously suppresses the SC phase, with both $T_c$ and $H_{c2}$ reaching zero at the critical concentration $x_c = 0.25$ \cite{doi:10.7566/JPSJ.85.094713}. These features can be confirmed in the $H$-$T$-$x$ phase diagram shown in Fig.\ 1 \cite{doi:10.7566/JPSJ.85.094713}. The AFM spin correlations govern the paramagnetic state after the SC order is completely suppressed at $x_c$. Specific heat, electrical resistivity, and magnetic susceptibility all exhibit NFL behavior at zero magnetic fields for $x_c$ \cite{doi:10.7566/JPSJ.85.094713,PhysRevB.99.054506}. The temperature dependence of the nuclear spin-lattice relaxation rate involves a broad peak at $\sim 2\ {\rm K}$, possibly arising from the nested AFM spin fluctuations \cite{PhysRevB.106.235152}. Furthermore, recent electrical resistivity measurements have revealed that the QCP responsible for the NFL behavior is consistently located at $H_{c2}\ (\neq 0)$ for $x < 0.25$ as well as at $H_{c2} = 0$ for $x_c$ \cite{PhysRevMaterials.8.L081801}. These Ni-doping effects are likely governed by changes in carrier density, similar to the mechanisms proposed for other doped compounds \cite{PhysRevLett.94.047001,PhysRevB.73.245109,PhysRevLett.105.126401,doi:10.7566/JPSJ.85.094713,PhysRevB.99.054506,PhysRevB.106.235152,PhysRevMaterials.8.L081801,PhysRevLett.109.186402,PhysRevB.97.045134,PhysRevB.105.115121,doi:10.1126/science.aaz4566,PhysRevB.109.125147}. However, it remains unclear how the doped Ni ions specifically influence the carrier density in CeCo$_{1-x}$Ni$_x$In$_5$.

In pristine CeCoIn$_5$, investigating the low-temperature and low-field behavior of $R_{\rm H}$ is hindered by the onset of superconductivity, which causes both electrical and Hall resistivities to vanish. In contrast, Ni-substituted CeCoIn$_5$ up to the critical concentration $x_c$ serves as an ideal platform for such studies, as NFL behavior persists down to zero field at $x_c$. This substituted system may offer crucial insights into the anomalous $R_{\rm H}$ features previously observed in CeCoIn$_5$, such as the significant reductions above $T_c$ and $H_{c2}$ \cite{doi:10.1143/JPSJ.73.5,PhysRevB.70.035113,doi:10.1143/JPSJ.76.024703,PhysRevLett.98.057001,PhysRevB.87.045102,doi:10.1126/science.aaz4566}. The origin of these anomalies remains a subject of intense debate due to the sensitivity of $R_{\rm H}$ to both magnetic correlations and Fermi surface topology. Given that CeCoIn$_5$ is considered a compensated metal characterized by a multi-band structure and strong correlations \cite{PhysRevB.64.212508,doi:10.1143/JPSJ.71.162,doi:10.1143/JPSJ.72.854}, it is of interest to determine how Ni doping modifies these anomalous Hall features.

Thus, we have measured the Hall resistivity of CeCo$_{1-x}$Ni$_x$In$_5$, including at the critical concentration of $x_c = 0.25$, to clarify the effects of Ni doping on the carrier density and the evolution of anomalous $R_{\rm H}$ behavior.

\section{Experiment Details}
\begin{figure}[!tpb]
\begin{center}
\includegraphics[keepaspectratio,width=0.45\textwidth]{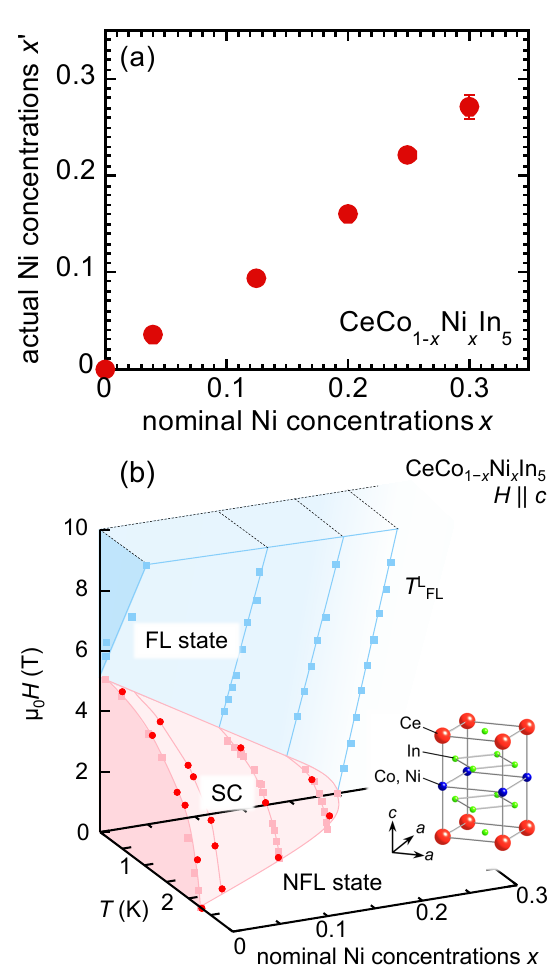}
\end{center}
\caption{
  (a) Comparison of nominal ($x$) and actual ($x'$) Ni concentrations measured by the inductively coupled plasma mass spectrometry measurements. The experimental errors in the actual Ni concentration are smaller than the size of the markers, except at $x=0.3$. (b) The $H$-$T$-$x$ phase diagram of CeCo$_{1-x}$Ni$_x$In$_5$ constructed using Hall resistivity data (circle markers), and previous electrical resistivity and magnetization data (square markers) for $H \parallel c$ \cite{PhysRevLett.91.246405,doi:10.7566/JPSJ.85.094713,PhysRevMaterials.8.L081801}. $T_{\rm FL}^{\rm L}$ denotes the crossover temperature between non-Fermi liquid (NFL) and Fermi liquid (FL) regions, estimated from the temperature dependence of the electrical resistivity $\rho(T)$ \cite{PhysRevMaterials.8.L081801}. The inset in Panel (b) shows the crystal structure of CeCo$_{1-x}$Ni$_x$In$_5$.
}
\end{figure}
Single crystals of CeCo$_{1-x}$Ni$_x$In$_5$ for $x \le 0.3$ were grown with an indium flux method, the details of which have been described in the previous report \cite{doi:10.7566/JPSJ.85.094713}. The chemical composition was analyzed using an inductively coupled plasma mass spectrometry (ICP-MS) system (7500cx, Agilent Technologies). Figure 1(a) and Table I show the relationship between the actual (measured) and nominal (pre-production stage) Ni concentrations. The actual Ni concentration $x'$ is defined as the molar ratio of Ni to the total constituent atoms in the chemical formula CeCo$_{1-x}$Ni$_x$In$_5$. It was confirmed that the molar ratios of Ce, (Co + Ni), and In are nearly 1, 1, and 5, respectively, and that the actual Ni concentrations closely match the nominal concentrations, as reported previously \cite{doi:10.7566/JPSJ.85.094713}. For simplicity, we use the nominal $x$ values throughout this article unless otherwise indicated to preserve the consistency with previous reports \cite{doi:10.7566/JPSJ.85.094713,PhysRevB.99.054506,PhysRevB.106.235152,PhysRevMaterials.8.L081801}. For current Hall resistivity measurements, we selected thin plate-shaped samples with a thickness $t$ along the $c$-axis of $\sim 0.1~{\rm mm}$ and a large $c$-plane surface (typical dimensions of $3 \times 2\ {\rm mm}^2$) to ensure the Hall component in the resistivity data. Note that the samples used for the Hall resistivity measurements and ICP-MS analysis were taken from the same plate; the latter was a tiny piece cut from the edge of the plate-shaped sample.

The electrical resistivity and Hall resistivity can be determined by calculating half of the sum and the difference of the voltages induced by the application of current, measured in the positive and negative directions of the magnetic field, respectively. The Hall coefficient $R_{\rm H}$ was determined by measuring the Hall voltage $V_{\rm H}$ along the tetragonal $a$-axis while applying a current $j$ along the $b$-axis (equivalent to the $a$-axis) and a magnetic field $H$ along the $c$-axis. We used the relation of $R_{\rm H}=tV_{\rm H}/j\mu_0H$, where $\mu_0$ is the vacuum permeability. This measurement was conducted using a commercial measurement system (PPMS, Quantum Design) within a temperature range of $T \ge 0.4$ K and a magnetic field range from $-9$ T to 9 T. The $R_{\rm H}$ sign was found to be negative across all temperatures, magnetic fields, and Ni concentrations investigated, consistent with the behavior observed in CeCoIn$_5$ \cite{doi:10.1143/JPSJ.73.5,PhysRevB.70.035113,doi:10.1143/JPSJ.76.024703,PhysRevLett.98.057001,PhysRevB.87.045102,doi:10.1126/science.aaz4566}.

In Fig.\ 1(b), we present the phase diagram for CeCo$_{1-x}$Ni$_x$In$_5$, illustrating the relationships among the magnetic field, temperature, and nominal Ni doping level. This diagram delineates the regions of NFL, Fermi liquid (FL), and SC states. It is confirmed that $T_c$ and $H_{c2}$ estimated from the Hall resistivity [indicated by circle markers in Fig.\ 1(b)] are nearly identical to those derived from previous electrical resistivity and magnetization measurements [represented by square markers in Fig.\ 1(b)] \cite{PhysRevLett.91.246405,doi:10.7566/JPSJ.85.094713,PhysRevMaterials.8.L081801}.

\section{Results}
\begin{figure}[!tpb]
\begin{center}
\includegraphics[keepaspectratio,width=0.45\textwidth]{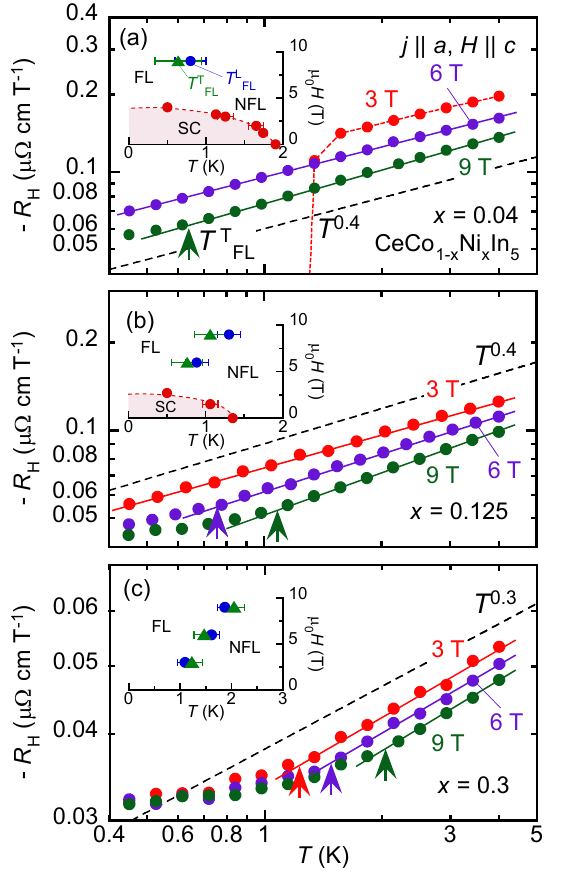}
\end{center}
\caption{
  The temperature dependence of the Hall coefficient $R_{\rm H}(T)$ for CeCo$_{1-x}$Ni$_x$In$_5$, with (a) $x=0.04$, (b) $x=0.125$, and (c) $x=0.3$. Note that logarithmic scales are used for both the vertical and horizontal axes. The solid lines illustrate the behavior of $-R_{\rm H}(T) \propto T^\alpha$, while the dashed lines serve as visual guides for the $T^\alpha$ slopes. The arrows indicate the NFL--FL crossover temperature $T_{\rm FL}^{\rm T}$ determined by a deviation from the relation of $-R_{\rm H}(T) \propto T^\alpha$. The insets show the magnetic field versus temperature phase diagrams for each Ni concentration, obtained by the electrical resistivity data (circle markers) and the Hall resistivity data (triangle markers).
}
\end{figure}

Figures 2(a)--2(c) display the temperature dependence of the Hall coefficient $R_{\rm H}(T)$ for CeCo$_{1-x}$Ni$_x$In$_5$, with $x=0.04,~0.125$, and 0.3, plotted on logarithmic scales for both the vertical and horizontal axes. Note that the negative sign is multiplied to $R_{\rm H}$ in the vertical axis of Figs.\ 2(a)--2(c) and subsequent plots in Figs. 3 and 4 because increasing the electron number generally enhances the $-R_{\rm H}(H)$ value, and the negative sign of $R_{\rm H}$ indicates that electron-like carriers govern the electrical transport properties in CeCo$_{1-x}$Ni$_x$In$_5$. $R_{\rm H}(T)$ exhibits the NFL behavior characterized by the relation of $-R_{\rm H}(T) \propto T^\alpha$ with $\alpha$ ranging from 0.3 to 0.5. However, as the temperature decreases, $R_{\rm H}(T)$ tends to be independent of temperature. This variation is suggested to be due to the formation of the FL states, a phenomenon frequently observed in the $R_{\rm H}(T)$ behavior of conventional metals \cite{pippard1989magnetoresistance}. The NFL-FL crossover temperatures $T_{\rm FL}^{\rm T}$ and $T_{\rm FL}^{\rm L}$ are determined from the deviation from the $T^\alpha$ behavior in $-R_{\rm H}$ and the onset of a $T^2$ dependence in the electrical resistivity $\rho(T)$, respectively. These temperatures were derived using a procedure based on the temperature derivative, similar to that reported previously \cite{PhysRevMaterials.8.L081801}. As shown in the $H$-$T$ phase diagrams plotted in the insets of Figs.\ 2(a)--2(c), $T_{\rm FL}^{\rm T}$ and $T_{\rm FL}^{\rm L}$ coincide well with each other within the error bars. This suggests that both the $R_{\rm H}(T)$ and $\rho(T)$ curves can capture the variation from the NFL state to the FL state. For the FL state at $x=0.3$, $R_{\rm H}(T)$ across the entire $H$ range exhibits a tendency to converge at low temperatures, as anticipated in the conventional FL state.

\begin{figure}[!tpb]
\begin{center}
\includegraphics[keepaspectratio,width=0.45\textwidth]{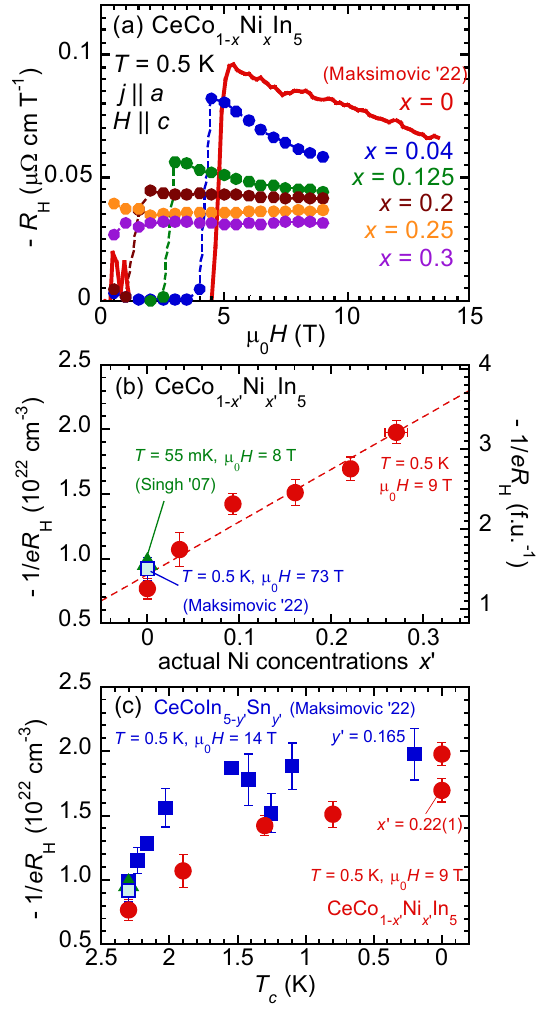}
\end{center}
\caption{
  (a) The dependence of the Hall coefficient $R_{\rm H}(H)$ on magnetic field and (b) the dependence of $-1/eR_{\rm H}$ ($e$: elementary charge) on the actual Ni concentration $x'$ at $\mu_0H = 9\ {\rm T}$ and $T = 0.5\ {\rm K}$ for CeCo$_{1-x}$Ni$_x$In$_5$, with $x \le 0.3$. In Panel (b), the left axis represents the values of $-1/eR_{\rm H}$ per unit volume, while the right axis displays the values per formula unit. The dashed line indicates the fitting result of $-1/eR_{\rm H}$ at $\mu_0H = 9\ {\rm T}$ and $T = 0.5\ {\rm K}$. (c) $-1/eR_{\rm H}$ at $T = 0.5~{\rm K}$ for CeCo$_{1-x'}$Ni$_{x'}$In$_5$ ($\mu_0H = 9~{\rm T}$) and CeCoIn$_{5-y'}$Sn$_{y'}$ ($\mu_0H = 14~{\rm T}$) \cite{doi:10.1126/science.aaz4566} plotted as a function of $T_c$, with actual concentration $x'$ and $y'$ as implicit parameters. The data for pure CeCoIn$_5$ are taken from Refs. \cite{doi:10.1126/science.aaz4566,PhysRevLett.98.057001} in Panels (a), (b), and (c).
}
\end{figure}
\begin{table}[!tbp]
\begin{center}
\caption{
  Nominal ($x$) and actual ($x'$) Ni concentrations, along with the values of $-1/eR_{\rm H}$ in CeCo$_{1-x}$Ni$_x$In$_5$. The temperature $T$ and magnetic field $\mu_0H$ at which $-1/eR_{\rm H}$ is estimated are also provided.
}
\begin{tabular}{ccccccccccc}
\hline \hline
$x$ && $x'$ && $T$ (K) && $\mu_0H$ (T) && $-1/eR_{\rm H}\ (10^{22} {\rm cm}^{-3})$ && \\
\hline
0 && 0 && 0.5 && 9 && 0.77(8)\\
0 && 0 && 0.5 && 73 && 0.92(9) \cite{doi:10.1126/science.aaz4566}\\
0 && 0 && 0.055 && 8 && 0.98 \cite{PhysRevLett.98.057001}\\
0.04 && 0.035(3) && 0.5 && 9 && 1.07(13)\\
0.125 && 0.093(4) && 0.5 && 9 && 1.42(8)\\
0.2 && 0.161(5) && 0.5 && 9 && 1.51(10)\\
0.25 && 0.221(7) && 0.5 && 9 && 1.70(9)\\
0.3 && 0.271(12) && 0.5 && 9 && 1.98(9)\\
\hline \hline
\end{tabular}
\end{center}
\end{table}

Figure 3(a) shows the magnetic field dependence of the Hall coefficient $R_{\rm H}(H)$ at $T = 0.5~{\rm K}$ for CeCo$_{1-x}$Ni$_x$In$_5$ with $x \le 0.3$.
The SC order can be confirmed by the zero value of $R_{\rm H}(H)$ for $H \le H_{c2}$.
In the normal state above $H_{c2}$, $-R_{\rm H}(H)$ increases as the magnetic field decreases toward $H_{c2}$ for $x \le 0.125$. In contrast, this enhancement of $-R_{\rm H}(H)$ around $H_{c2}$ is significantly diminished for $x \ge 0.2$. This trend indicates that Ni doping effectively quenches the anomalous enhancement of $R_{\rm H}$ near $H_{c2}$, suggesting a suppression of the underlying mechanism.

The carrier density $n$ is expected to be expressed by $n = n_e - n_h = -1/eR_{\rm H}$ in sufficiently high magnetic fields, where $n_e$ and $n_h$ represent the concentrations of electrons and holes, respectively \cite{pippard1989magnetoresistance}. In Fig.\ 3(b), we illustrate the dependence of actual Ni concentration $x'$ on $-1/eR_{\rm H}$ at $T = 0.5\ {\rm K}$ and $\mu_0H = 9\ {\rm T}$. The monotonic increase in $-1/eR_{\rm H}$ with increasing Ni concentration suggests that the carrier density of negative charge rises as a result of electron doping associated with the substitution of Ni. The value of $-1/eR_{\rm H}$ increases from 1.2(1) to 3.2(1) number/f.u. with increasing $x'$ from 0 to 0.27(1), resulting in a rate of change of $\partial (-1/eR_{\rm H})/\partial x'=6.5(6)\ {\rm number/(f.u.\ Ni)}$ [the dashed line in Fig. 3(b)]. This rate is significantly higher than the expected rate of $1.0\ {\rm number/(f.u.\ Ni)}$ based on the assumption that a conduction electron is supplied solely from an excess electron of the doped Ni ion in CeCoIn$_5$.

\begin{figure*}[!tpb]
\begin{center}
\includegraphics[keepaspectratio,width=0.95\textwidth]{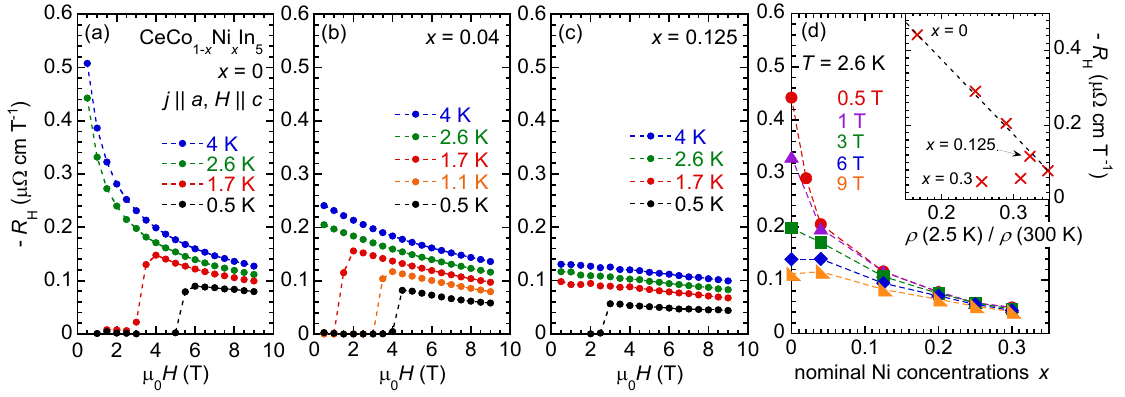}
\end{center}
\caption{
  Magnetic field dependence of the Hall coefficient $-R_{\rm H}(H)$ for CeCo$_{1-x}$Ni$_x$In$_5$, with (a) $x = 0$, (b) $x = 0.04$, and (c) $x = 0.125$. Panel (d) shows $-R_{\rm H}$ at $T = 2.6~{\rm K}$ and $\mu_0H = 0.5$, 1, 3, 6, and 9 T for CeCo$_{1-x}$Ni$_x$In$_5$, with $x \le 0.3$, plotted as a function of Ni concentration $x$. The inset in Panel (d) illustrates the $-R_{\rm H}$ values for $T = 2.6\ {\rm K}$ and $\mu_0H = 0.5\ {\rm T}$ plotted as a function of the inverse of the residual resistivity ratio $\rho(2.5\ {\rm K})/\rho(300\ {\rm K})$, obtained by $x$ as an implicit parameter.
}
\end{figure*}

For $x \ge 0.2$, $R_{\rm H}$ is field-independent in the paramagnetic state at $T = 0.5\ {\rm K}$, indicating that the carrier density can be reliably estimated [Fig.\ 3(a)]. However, for $x \le 0.125$, the non-linear field dependence is observed near $H_{c2}$ in $R_{\rm H}(H)$, which tends to weaken as the magnetic field increases up to 9 T at 0.5 K [Fig.\ 3(a)]. This enhancement in $-R_{\rm H}$ cannot be explained by the simple single band model; it is likely due to the specific band structure and/or strong spin correlations. In fact, band structure calculations and de Haas--van Alphen measurements suggest that CeCoIn$_5$ is a compensated multiband metal with several bands crossing the Fermi level \cite{PhysRevB.64.212508,doi:10.1143/JPSJ.71.162,doi:10.1143/JPSJ.72.854}. We have used the $R_{\rm H}$ values at 9 T, the maximum magnetic field applied in the current measurements, to estimate the carrier density. However, it is important to consider whether this magnetic field strength is sufficient. We confirmed that pure CeCoIn$_5$ exhibits the strongest field dependence of $R_{\rm H}$ above $H_{c2}$ among all Ni compositions [Fig.\ 3(a)]. In this compound, $R_{\rm H}(H)$ has been measured up to 73 T at $0.5\ {\rm K}$, and it remains constant above 15 T \cite{doi:10.1126/science.aaz4566}. This suggests that the nontrivial effects causing the non-linear $R_{\rm H}(H)$ behavior are likely negligible in this high magnetic field region. As shown in Fig.\ 3(b), the carrier density estimated from the $R_{\rm H}$ data at 73 T and 0.5 K closely matches both our data at 9 T and 0.5 K, and the other low-temperature data at $55\ {\rm mK}$ and 8 T \cite{PhysRevLett.98.057001}, within experimental accuracy. Furthermore, the differences in the carrier density values among the Hall resistivity experiments in CeCoIn$_5$ are found to be significantly smaller than the changes in the carrier density induced by the Ni doping [Fig.\ 3(b)]. Even when we adopt the carrier density defined at the high fields for CeCoIn$_5$ \cite{doi:10.1126/science.aaz4566} and those obtained from the current measurement only for $x \ge 0.2$, where $R_{\rm H}$ is independent of $H$ [Fig.\ 3(a)], we evaluate the rate which is nearly the same as the estimate obtained using the entire current data set.

CeCoIn$_5$ exhibits an enhancement of $-R_{\rm H}$ above $T_c\ (=2.3\ {\rm K})$ at low magnetic fields, as well as at low temperatures and magnetic fields near $H_{c2}$ \cite{doi:10.1143/JPSJ.73.5,PhysRevB.70.035113,doi:10.1143/JPSJ.76.024703}. To clarify where the enhancement of $-R_{\rm H}$ is most pronounced in the $H$-$T$-$x$ space, we plot the $R_{\rm H}(H)$ curves for $x = 0$, 0.04, and 0.125 in Figs. 4(a)--4(c), respectively. For pure CeCoIn$_5$, the evolution of $-R_{\rm H}(H)$ as the magnetic field decreases toward zero for $T > T_c$ (2.3 K) is more pronounced than that observed for $H > H_{c2}$ at $T = 0.5\ {\rm K}$ [Fig. 4(a)]. This enhanced behavior in $-R_{\rm H}(H)$ is significantly suppressed at $x = 0.04$ [Fig. 4(b)] and subsequently changes into an approximately linear field dependence in $R_{\rm H}(H)$ for $x \ge 0.125$ [Fig. 4(c)]. To emphasize the trends above induced by the Ni substitution, we summarized the $x$ dependence of $R_{\rm H}$ at $T = 2.6~{\rm K}$ in Fig. 4(d). It is evident that $-R_{\rm H}$ significantly decreases as the magnetic field or Ni concentration initially increases, while its variation is relaxed at high fields or elevated Ni concentrations. As for weak magnetic fields and temperatures above $T_c$ in CeCoIn$_5$, previous experiments revealed that $-R_{\rm H}$ exhibits a maximum at $\sim 4$ K, which is similar to the behavior observed in our measurements [Fig.\ 4(a)] \cite{doi:10.1143/JPSJ.73.5,PhysRevB.70.035113,doi:10.1143/JPSJ.76.024703}.

\begin{figure}[!tpb]
\begin{center}
\includegraphics[keepaspectratio,width=0.45\textwidth]{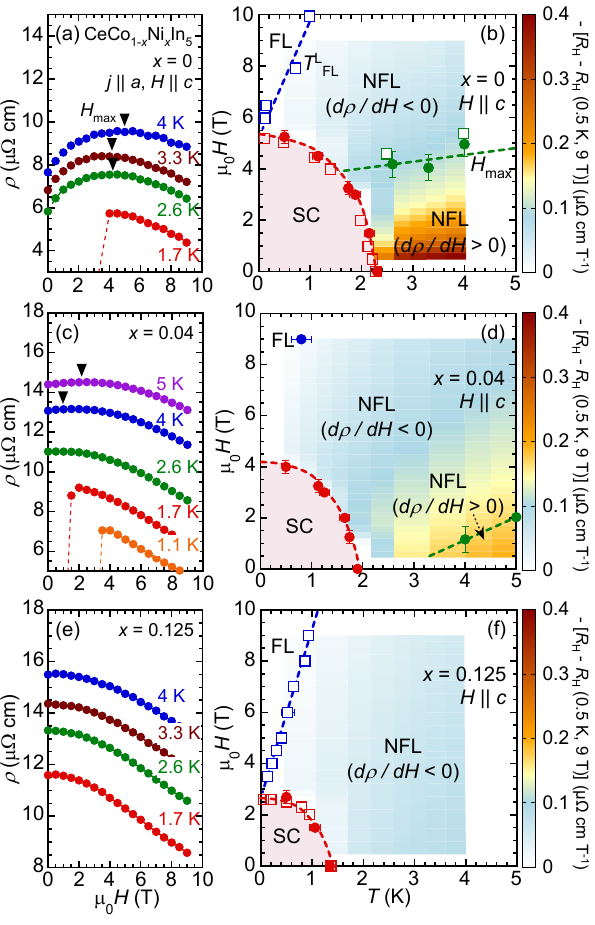}
\end{center}
\caption{
  The magnetic field dependence of the electrical resistivity $\rho(H)$ for CeCo$_{1-x}$Ni$_x$In$_5$, with (a) $x = 0$, (c) $x = 0.04$, and (e) $x = 0.125$. The arrows in Panels (a) and (c) indicate the field $H_{\rm max}$ at which $\rho(H)$ reaches a maximum. The $H$-$T$ phase diagram of CeCo$_{1-x}$Ni$_x$In$_5$ for (b) $x = 0$, (d) $x = 0.04$, and (f) $x = 0.125$, constructed using current electrical resistivity data (closed circle markers) and the previous electrical resistivity data (open square markers) \cite{PhysRevLett.91.246405,PhysRevMaterials.8.L081801}. $T_{\rm FL}^{\rm L}$ is defined as the temperature where the electrical resistivity deviates from the $T^2$ behavior characteristic of the FL state. In Panels (b), (d), and (f), the color plots represent the magnitude of the Hall coefficient $-R_{\rm H}$, from which the value at 0.5 K and 9 T has been subtracted, and the dashed lines are guides for the eyes.
}
\end{figure}
Here, we compare the anomalous enhancement of $-R_{\rm H}$ with certain characteristics observed in the electrical resistivity. Figures 5(a), 5(c), and 5(e) illustrate the magnetic field dependence of the electrical resistivity $\rho(H)$ for CeCo$_{1-x}$Ni$_x$In$_5$, with $x = 0$, 0.04, and 0.125, respectively, as obtained from the current measurements. For $x = 0$, $\rho(H)$ exhibits a maximum at the field $H_{\rm max}$ for $T > T_c$ [Fig.\ 5(a)]. This feature has also been observed in the previous electrical resistivity measurements \cite{PhysRevLett.91.246405}, and both sets of the measurements provide nearly identical positions of $H_{\rm max}$ in the $H$-$T$ phase diagram [Fig.\ 5(b)]. Note that in the $H$-$T$ phase diagram, $H_{\rm max}$ defines the boundary that separates the regions where $d\rho/dH > 0$ and $d\rho/dH < 0$ in the NFL state \cite{PhysRevLett.91.246405}. It is found that the $H_{\rm max}$ boundary shifts to lower magnetic fields for $x = 0.04$ [Figs.\ 5(c) and 5(d)], and it subsequently disappears entirely for $x = 0.125$ [Figs.\ 5(e) and 5(f)]. Similar results have been obtained in our recent electrical resistivity measurements, in which $\rho(H)$ decreases monotonically with $H$ without exhibiting a maximum for $x \ge 0.05$ \cite{PhysRevMaterials.8.L081801}. Interestingly, the suppression of the divergent-like behavior in $-R_{\rm H}$ above $T_c$ at low magnetic fields is likely coupled with the disappearance of the NFL region characterized by the condition of $d\rho/dH > 0$ because both features occur sensitively with the doping of small amounts of Ni ions. In the NFL region where $d\rho/dH > 0$ for $x = 0$ and $0.04$, $-R_{\rm H}(H)$ exhibits a nonlinear decrease as the magnetic field increases [Figs.\ 4(a) and 4(b)]. However, the $-R_{\rm H}(H)$ curve changes to a weak linear dependence on the magnetic field for $x \ge 0.125$ [Fig.\ 4(c)], where the region with $d\rho/dH > 0$ disappears in the $H$-$T$ phase diagram [Fig.\ 5(f)]. As shown in Figs. 5(b), 5(d), and 5(f), the correspondence between the regions where $-R_{\rm H}$ is significantly enhanced and where $d\rho/dH > 0$ holds true can also be confirmed by the color maps of the $-R_{\rm H}$ values. The total Hall coefficient is generally expressed as the sum of a normal term, which primarily reflects the carrier density, and an anomalous term, which reflects other effects \cite{RevModPhys.82.1539,Nair01102012}. As discussed in relation to Fig.\ 3(b), we assume that the $R_{\rm H}$ value at 0.5 K and 9 T represents the normal term. To separate the variations in the carrier density due to Ni substitution, the $-R_{\rm H}$ values in the color maps represent the total Hall coefficient minus the $R_{\rm H}$ value at 0.5 K and 9 T.

It is unclear why the enhancement of $-R_{\rm H}$ is significantly suppressed by even minor amounts of Ni substitution. A possible explanation is the alteration of the electronic states resulting from electron doping, as indicated by the current measurements shown in Fig.\ 3(b). Alternatively, one may consider that the effects of impurity scattering enhanced due to doping may contribute to the reduction of $-R_{\rm H}$; specifically, Ni doping decreases the residual resistivity ratio $\rho(300\ {\rm K})/\rho(2.5\ {\rm K})$ from 6.0 ($x=0$) to 3.1 ($x=12.5$), and this trend appears to correlate with the observed significant reduction in $-R_{\rm H}$ at weak magnetic fields and at up to $x = 0.2$ [see the inset in Fig.\ 4(d)]. However, in the La-substituted system, where the La$^{3+}$ ion lacks 4$f$ electrons and is primarily expected to introduce defects at magnetic sites, the residual resistivity $\rho(T\rightarrow 0)$ increases by approximately a factor of five as the La concentration rises from 0\% to 6\% \cite{doi:10.1126/science.aaz4566}. Despite this significant increase in residual resistivity caused by the impurity scattering, the magnitude of $-R_{\rm H}$ at $T = 2.5\ {\rm K}$ and $\mu_0H = 0.1\ {\rm T}$ remains nearly unchanged up to 6\% La substitution \cite{doi:10.1126/science.aaz4566}. Furthermore, $H_{\rm max}$ also remains nearly constant for La concentrations up to 10\% \cite{paglione2005quantum}. This contrasting trend between Ni- and La-doped compounds suggests that the enhancement of $-R_{\rm H}$ under the condition of $d\rho/dH > 0$ is not primarily influenced by impurity scattering. Further insights may be obtained from the behavior of the Ni-substituted system, in which the peak in the magnetic-field dependence of electrical resistivity does not simply broaden while maintaining a fixed maximum position $H_{\rm max}$; rather, it shifts toward lower magnetic fields as $x$ increases, accompanied by a shrinkage in the region where $-R_{\rm H}$ increases. This suggests that Ni doping not only induces local impurity effects but also modifies the energy scale of the coherent electronic state.

\section{Discussion}
\subsection{The effect of electron doping due to Ni substitution}
We observed that $-1/eR_{\rm H}$ at 0.5 K and 9 T continuously increases with increasing $x$ [Fig.\ 3(b)]. This increase is attributed to the rise in conduction electrons due to the Ni substitutions, as expected from the electron configurations of Co [$3d^7$ $4s^2$] and Ni [$3d^8$ $4s^2$]. However, the rate of change in the carrier density caused by doping Ni ion, $\partial n/\partial x$, is approximately 6.5 times greater than unity, which is expected when assuming that an excess $3d$ electron in the doped Ni ion solely contributes to the conduction electrons in Ni-doped CeCoIn$_5$. The observed enhancement in the electron number suggests that doping Ni into CeCoIn$_5$ not only increases the number of conduction electrons but also yields some modifications of the band structure around the Fermi energy. Note that the effective magnetic moment for $x \le 0.3$, estimated from the temperature dependence of magnetic susceptibility, exhibits little variation with $x$ and closely matches the value estimated from the $J = 5/2$ multiplet of the Ce$^{3+}$ ion \cite{doi:10.7566/JPSJ.85.094713}. This suggests that the Ni 3$d$ electrons do not significantly contribute to the magnetic properties through the formation of localized moments in Ni-substituted CeCoIn$_5$. In this regard, it is interesting to compare the variations in carrier density for Ni-doped alloys with those for Sn substitutions, the latter involving similar doping conditions concerning the electron configurations: In [$5s^2$ $5p^1$] and Sn [$5s^2$ $5p^2$]. Figure 3(c) shows the magnitude of $-1/eR_{\rm H}$ (interpreted as $n$) at $T = 0.5\ {\rm K}$ as a function of $T_c$, obtained by using actual Ni concentration $x'$ in CeCo$_{1-x'}$Ni$_{x'}$In$_5$ and the actual Sn concentration $y'$ in CeCoIn$_{5-y'}$Sn$_{y'}$ as implicit parameters \cite{doi:10.1126/science.aaz4566}. Here, we express the chemical formula as CeCoIn$_{5-y'}$Sn$_{y'}$ instead of CeCo(In$_{1-y'}$Sn$_{y'}$)$_5$ to ensure consistency in the amount of substitution with CeCo$_{1-x'}$Ni$_{x'}$In$_5$. The magnitude of $-1/eR_{\rm H}$ at the critical concentrations, characterized by the $T_c \to 0$ condition, is comparable between these alloys. Therefore, it is considered that both Ni and Sn ions significantly influence the electronic states, exceeding the effect of merely increasing their number. However, $-1/eR_{\rm H}$ exhibits different curves with decreasing $T_c$ between these alloys. For the Ni-doped alloys, $-1/eR_{\rm H}$ increases with an approximately linear dependence on $T_c$, whereas in the Sn-substituted alloys, it is steeply enhanced over a small range of $y'$ with a rate of change $\partial n/\partial y' =27(3)$ number/(f.u. Sn), followed by a tendency of saturation around $T_c \sim 2\ {\rm K}$ \cite{doi:10.1126/science.aaz4566}. These discrepancies in the $-1/eR_{\rm H}$ curves are likely caused by the differences in the effects of ion doping into CeCoIn$_5$. More specifically, the Fermi surfaces and charge compensation conditions in CeCoIn$_5$ are suggested to be highly sensitive to the type of doped ions \cite{PhysRevB.64.212508,doi:10.1143/JPSJ.71.162,doi:10.1143/JPSJ.72.854}, because the orbital angular momentum of the excess electron in Ni- and Sn-doped CeCoIn$_5$ exhibits $d$-like and $p$-like characteristics, respectively, and the doped ions occupy the different crystallographic sites between these alloys. Alternatively, this may also be related to the instability associated with the recently proposed quantum phase transition in Sn-substituted CeCoIn$_5$, which involves a localized-to-delocalized crossover of Ce 4$f$ electrons without magnetic ordering \cite{doi:10.1126/science.aaz4566}. We expect that the photoemission spectroscopy investigation would provide valuable information on this issue.

The relationship between the Fermi surface and quantum criticality is a subject of intense research interest. In AFM compounds such as YbRh$_2$Si$_2$ and Cr, the evolution of the Fermi surface volume across the QCP has been investigated through Hall resistivity measurements. The results suggest that the QCP in YbRh$_2$Si$_2$ arises from the so-called Kondo breakdown mechanism, whereas in Cr, it is associated with the spin-density-wave mechanism \cite{paschen2004hall,doi:10.1073/pnas.1009202107,PhysRevLett.92.187201,doi:10.1073/pnas.1005036107}. In contrast, such an investigation appears challenging for Ni-substituted CeCoIn$_5$ because the QCP is known to be located very close to $H_{c2}(0)$ for $x \le 0.25$ \cite{PhysRevMaterials.8.L081801}, at which the Hall effect is significantly influenced by the SC phase. Furthermore, $-R_{\rm H}$ does not show a discontinuous change with $x$, but instead exhibits a linear increase up to $x\sim 0.30$ [Fig. 3(b)], although the quantum critical behavior is significantly weakened at $x=0.30$ after the SC order completely disappears, reflecting that Ni doping suppresses superconductivity without inducing an AFM phase \cite{doi:10.7566/JPSJ.85.094713}. Consequently, based on the current data, it is difficult to conclude whether the underlying mechanism of the quantum criticality corresponds to Fermi surface reconstruction (the Kondo destruction scenario) or scattering anisotropy associated with magnetic ordering (the spin-density-wave scenario). Further measurements that tune the system across a magnetic QCP, such as applying pressure or co-doping with other ions, are required to distinguish between these two scenarios.

\subsection{The origin of the enhancement in the Hall coefficient}
The current investigation revealed that at low Ni doping levels in CeCo$_{1-x}$Ni$_x$In$_5$, the magnitude of $-R_{\rm H}$ is strongly enhanced at magnetic fields near the superconducting upper critical field $H_{c2}$ and in the low-field region above the superconducting transition temperature $T_c$. However, these enhancements are diminished by further increases in Ni concentration. In general, the Hall coefficient reflects not only the carrier density but also the shape of the Fermi surface resulting from complex band structures and magnetic correlations. Therefore, identifying the origin of its anomalous behavior is challenging in many strongly correlated electron systems.

This is also the case for CeCoIn$_5$. The weak dependence of $-R_{\rm H}$ on temperature and magnetic field is observed in the paramagnetic state of the non-$4f$ isostructural compound LaCoIn$_5$ at low temperatures; however, its magnitude is approximately one-fiftieth of that in CeCoIn$_5$ at $H \to 0$ and $T = 2.5\ {\rm K}$ \cite{PhysRevB.70.035113,doi:10.1143/JPSJ.76.024703}. This comparison between CeCoIn$_5$ and LaCoIn$_5$ suggests that the anomalous enhancements of $-R_{\rm H}$ in CeCoIn$_5$ are primarily caused by the involvement of Ce 4$f$ electrons. Here, we discuss the effects of the effective charge density of renormalized quasiparticles and/or magnetic correlations on the increase in $-R_{\rm H}$ induced by Ce 4$f$ electrons. This discussion is based on comparisons with previous interpretations of the enhancement of $R_{\rm H}$ in CeCoIn$_5$ \cite{PhysRevB.70.035113,doi:10.1143/JPSJ.76.024703,PhysRevB.87.045102}, as well as physical quantities observed in Ni-substituted systems.

Thus far, two scenarios have been proposed to explain the enhancement of $-R_{\rm H}$ in CeCoIn$_5$. The first is the two-fluid Kondo model, which attributes the enhancement of $-R_{\rm H}$ to a reduction in the effective carrier density associated with the formation of Kondo singlets \cite{PhysRevB.70.035113, PhysRevB.87.045102}. Indeed, at low magnetic fields, the temperature at which $-R_{\rm H}$ begins to increase ($T \sim 50\ {\rm K}$) with decreasing temperature coincides with the Kondo lattice coherence temperature derived from this model \cite{PhysRevLett.92.016401}. In this context, introducing Ni ions as impurities is expected to disrupt the periodic potential of the Kondo lattice and hinders the formation of the coherent Kondo state, offering a straightforward explanation for the suppression of $-R_{\rm H}$ with Ni doping [Fig.\ 4(d)]. However, the electrical resistivity measurements revealed the opposite trend; Ni substitution increases the coherence temperature $T_{\rm coh}$ defined by the broad peak in the temperature dependence of electrical resistivity, from 46(1) K at $x=0$ to 60(1) K at $x=0.25$ \cite{doi:10.7566/JPSJ.85.094713}. Along with this increase, the electronic specific heat coefficient $\gamma$ at $\mu_0H = 7\ {\rm T}$ ($H \parallel c$) also decreases from $0.9(1)\ {\rm J/(K^2\cdot mol)}$ at $x=0$ to $0.36(6)\ {\rm J/(K^2\cdot mol)}$ at $x=0.25$ \cite{PhysRevLett.91.257001,PhysRevB.99.054506}. Since $\gamma$ is inversely proportional to the Kondo temperature $T_{\rm K}$, and $T_{\rm coh}$ generally correlates with $T_{\rm K}$ in most heavy-fermion systems \cite{RevModPhys.56.755,yang2008scaling}, these experimental results suggest that the coherent Kondo state does not become unstable with increasing Ni concentration. To further investigate the validity of this interpretation, it is informative to compare these features with those observed in Cd-, Hg-, or Zn-substituted systems, which are ascribed to possible hole-doped systems where $T_{\rm coh}$ is expected to decrease with ion doping \cite{doi:10.7566/JPSJ.83.033706,PhysRevLett.109.186402}.

In the second scenario, the increase in $-R_{\rm H}$ above $T_c$ at low magnetic fields is attributed to AFM quantum critical fluctuations (QCFs) arising from anisotropic scattering times at hot spots on the Fermi surface. This interpretation aligns with previous Hall resistivity studies of CeCoIn$_5$ \cite{doi:10.1143/JPSJ.76.024703}. This feature may be associated with the NFL behavior corresponding to a linear $T$-dependence of resistivity at zero magnetic field. Interestingly, similar signatures indicative of a possible NFL state in the low-field region have been observed in the Nernst coefficient of CeCoIn$_5$, an off-diagonal transport quantity analogous to the Hall resistivity \cite{PhysRevLett.99.147005}. Although these indications are limited to transport quantities, the anomalies consistently appear in the same temperature and magnetic field regions, suggesting a common origin. However, several issues remain unresolved in this scenario. It is unclear why the QCFs persist significantly up to $\sim 4\ {\rm K}$ or higher, whereas they often weaken around 1 K due to thermal effects in heavy-fermion compounds exhibiting quantum criticality. In addition, magnetic order associated with these possible QCFs emerging in the region where $d\rho/dH > 0$ has not yet been observed in CeCoIn$_5$ and its doped alloys. For Ni-doped CeCoIn$_5$, the NFL behavior in the temperature dependence of electrical resistivity persists up to $x=0.25$ \cite{doi:10.7566/JPSJ.85.094713}, whereas the enhancement in $-R_{\rm H}$ rapidly diminishes for $x \ge 0.125$ [Fig.\ 4(d)]. A direct observation of QCFs using inelastic neutron scattering would clarify whether the increase in \(-R_{\rm H}\) in the $d\rho/dH > 0$ regime is attributable to the QCFs.

We compare our Hall effect results with existing experimental data in the $H$-$T$ phase diagram. We find that the region where $-R_{\rm H}(H)$ is strongly enhanced (at $H \sim 0$ and $T > T_c$) does not overlap with the quantum critical region [$\sim H_{c2}(0)$] identified by specific heat and electrical resistivity measurements in CeCoIn$_5$ \cite{PhysRevLett.91.246405, PhysRevLett.91.257001}. In CeCoIn$_5$, the QCP at $H_{c2}(0)$ is characterized by a $-\ln T$ divergence in specific heat and a pronounced enhancement of the $A$ coefficient in electrical resistivity ($\rho \propto T^2$) \cite{PhysRevLett.91.246405, PhysRevLett.91.257001}. 
The NFL--FL crossover temperature $T^{\rm L}_{\rm FL}$ and the characteristic field $H_{\rm max}$ both extrapolate toward $H_{c2}(0)$ as $T \to 0$ [Fig. 5(b)], suggesting their association with the QCP \cite{PhysRevLett.91.246405}. However, the Ni-doped samples reveal a different trend. While the QCP persists at $H_{c2}(0)$ with increasing $x$ up to $0.25$, as evidenced by the divergence of the $A$ coefficient \cite{PhysRevMaterials.8.L081801}, even slight Ni substitution markedly shifts the $H_{\rm max}$ boundary toward lower fields, leading to its disappearance for $x \ge 0.05$ [Figs. 5(b), 5(d), and 5(f)]. 
This decoupling indicates that the $H_{\rm max}$ boundary is not intrinsically tied to the QCP at $H_{c2}(0)$. Intriguingly, the strong enhancement of $-R_{\rm H}$ at $H \sim 0$ and $T > T_c$ is confined to $H < H_{\rm max}$ within a narrow Ni concentration range [Fig. 4(d)], where $\rho$ shows no significant anomalies beyond $d\rho/dH > 0$ and $\rho \propto T$. Conversely, near $H_{c2}$ at 0.5 K, the magnitude of $-R_{\rm H}$ remains modest for all $x \le 0.25$, despite clear QCFs appearing in $\rho$ \cite{PhysRevMaterials.8.L081801}. These results suggest that the anomalous transport properties near $H_{c2}$ and those in the low-field region originate from distinct mechanisms: the former is primarily coupled with $\rho$, while the latter is coupled with $R_{\rm H}$. Nevertheless, the NFL–FL crossover in the $d\rho/dH < 0$ region is detectable in both $\rho(T)$ and $R_{\rm H}(T)$ (Fig. 2), suggesting a partial coupling between QCFs and these transport quantities.

\section{conclusion}
We investigated the effect of Ni substitution on the carrier density and anomalous electrical transport properties in CeCo$_{1-x}$Ni$_x$In$_5$, with $x \leq 0.3$, using Hall resistivity measurements. The Hall coefficient $R_{\rm H}$ exhibits a characteristic temperature dependence associated with the NFL--FL crossover in the normal state; a power-law temperature dependence indicative of the NFL behavior changes into a nearly temperature-independent behavior at low temperatures, corresponding to the formation of the FL state. The carrier density estimated from $-1/eR_{\rm H}$ at $T = 0.5\ {\rm K}$ in sufficiently high magnetic fields increases linearly with Ni doping, reflecting an increase in the electron carrier concentration provided by the doped Ni ions.

It was also found that in pure and lightly Ni doped CeCoIn$_5$, $-R_{\rm H}(H)$ is significantly enhanced in the high-field regime above $H_{c2}$ at low temperatures, as well as in the low-field regime above $T_{\rm c}$. However, the enhancement of $-R_{\rm H}(H)$ in both regimes is suppressed by Ni doping. In particular, in the low-field region above $T_{\rm c}$ where the condition $d\rho/dH > 0$ is simultaneously satisfied, this feature is rapidly diminished as the Ni concentration increases beyond $x = 0.05$. We discuss the possible origins of this anomalous behavior in $-R_{\rm H}(H)$ from the viewpoints of the Kondo effect and QCFs.
\vspace{20pt}

\begin{acknowledgments}
We express gratitude to H. Sakai and I. Kawasaki for helpful discussions. A part of this research was carried out as joint research at the Institute for Materials Research, Tohoku University, and was supported by the Japan Society for the Promotion of Science KAKENHI Grants No. 17K05529, No. 20K03852, and No. 23K03315. We thank to the technical assistance of Open Facility Center for Research, Ibaraki University. R.K. was supported by a research grant from the Murata Science and Education Foundation and by JST SPRING, Grant No. JPMJSP2161.
\end{acknowledgments}

\section*{Data availability}
The data that support the findings of this article are openly available \cite{Koizumi2026data}.


%
\end{document}